\documentstyle[12pt,preprint,tighten,prd,aps]{revtex}
\begin{document}
\draft
\title{Hilbert space of wormholes}
\author{Luis J. Garay\thanks{E--mail address: {\bf
Garay@cc.csic.es}.}}
\address{ESA IUE Observatory,
\thanks{Affiliated to the
Astrophysics Division, Space Science Department, ESTEC.}
P. O. Box 50727, E--28080 Madrid, Spain\\
and\\
Instituto de Optica, Consejo Superior de Investigaciones
Cient\'{\i}ficas\\ Serrano 121, E--28006 Madrid, Spain}
\date{\today}
\maketitle
\begin{abstract}
Wormhole boundary conditions for the Wheeler--DeWitt equation
can be  derived from the path integral formulation. It is
proposed that the wormhole wave function must be square
integrable in the maximal analytic extension of minisuperspace.
Quantum wormholes can be invested with a Hilbert space
structure, the inner product being naturally induced by the
minisuperspace metric, in which the Wheeler--DeWitt operator is
essentially self--adjoint. This provides us  with a kind of
probabilistic interpretation.  In particular, giant wormholes
will give extremely small contributions to any wormhole state.
We also study the whole spectrum of the Wheeler--DeWitt
operator and its role in the calculation of Green's functions
and effective low energy interactions.
\end{abstract}
\narrowtext


\section{Introduction}
\label{sec:intro}

Wormholes may affect the constants of nature through  low
energy effective interactions \cite{ha88,co88}. In particular
they may drive the cosmological constant to zero \cite{co88}
and select general relativity as the low energy theory of
gravity among Jordan--Brans--Dicke theories \cite{gg92}. In
order to determine which are these interactions, it is
necessary to have a well defined Hilbert space structure for
wormholes \cite{ha88}.

The quantum wormhole wave function is given by the path
integral over all possible asymptotically Euclidean spacetimes
and over all matter fields defined on them whose energy
momentum tensor vanishes at infinity; it is  typically labelled
by the asymptotic matter field configuration. Formally, this
wave function satisfies the Wheeler--DeWitt equation  and the
diffeomorphism constraints. Therefore, in order to find
wormhole wave functions, one can equivalently solve these
equations with appropriate boundary conditions which can be
derived from those for the path integral formulation
\cite{hp90,cg91}. Since spacetime is asymptotically Euclidean,
the wave function will be exponentially damped for large
three--geometries. On the other hand, for small
three--geometries no singularities are expected and therefore,
the wave function should be regular at these configurations in
some sense, so that the absence of singularities be properly
reflected \cite{hp90}. At least in the minisuperspace models
studied so  far, the wave functions calculated via path
integrals are regular at every configuration in minisuperspace
and, in particular, at the configurations that represent a zero
volume three--geometry \cite{ga91}. This expresses the fact
that although the three--geometry degenerates, the
four--geometries under consideration are perfectly regular. The
existence, in these circumstances,  of three--geometries with
zero volume is due to the slicing procedure which has been
carried out in spacetime. Therefore, the wormhole wave function
must be regular at any field configuration and any
three--geometry.  Finally, as it happens in ordinary quantum
mechanics, the wave function must vanish at infinite field
values, since these configurations cannot dominate the wave
function. All these conditions suggest that the wave function
must be square integrable in superspace.

Due to the well known difficulties in studying superspace, we
shall concentrate in  minisuperspace. Although the
generalization of the results thus obtained to full superspace
is not straightforward at all, it may be expected that some
qualitative results will still hold. In minisuperspace, the
Euclidean action takes the form
\[
I=\int_0^\infty d\tau N\left(\frac{1}{N^2}f_{\mu\nu}\dot
q^\mu\dot q^\nu+V(q)\right) \ ,
\]
where $q^\mu$ ($\mu=1,2,\ldots,n$)  are the degrees of freedom
that represent the three--geometry and the matter fields; $N$
is the lapse function which ensures the invariance under time
reparametrizations; $f_{\mu\nu}$ is the metric in
minisuperspace, so that  the line element in minisuperspace  is
\[
d{\cal S}^2=\frac{1}{N}f_{\mu\nu}dq^\mu dq^\nu \ ;
\]
and $V(q)$ is the Wheeler--DeWitt potential, which does not
contain any time derivatives $\dot q^\mu$. For each choice of
the lapse function, there is a different metric in
minisuperspace, all these metrics being  related by conformal
transformations. Therefore, for each of these choices, the
quantization process will be different. However we will argue
that physical results are independent of this gauge choice.
Although the minisuperspace metric $f_{\mu\nu}$ has hyperbolic
signature $(-+\cdots +)$, this does not mean  that there exists
a time variable in minisuperspace since superspace posseses no
timelike Killing  vectors. In fact, wormhole boundary
conditions do not give any privilige at all to any of the
minisuperspace variables.

 In this paper, we analyze the wormhole boundary conditions and
give them a precise meaning (Sec.~\ref{sec:bcs}).
Sec.~\ref{sec:hs}  is devoted to the study of the wormhole
Hilbert space.  We analyze the expression for the low energy
effective interaction caused by wormholes in view of this
structure. Also in this section, the Green's functions of the
Wheeler--DeWitt operator are considered in relation to the full
spectrum of this operator. The minisuperspace models obtained
by minimally or conformally coupling a massless homogeneous
scalar field to a homogeneous and isotropic spacetime
illustrate the Hilbert space structure of wormholes
(Sec.~\ref{sec:mss}). We summarize and conclude in
Sec.~\ref{sec:con}.


\section{Boundary conditions}
\label{sec:bcs}

The square integrability wormhole boundary condition in
minisuperspace  ensures that the wormhole wave function is
damped for large values of the three--volume and the matter
fields. It also implies regularity of the wave function  in the
whole minisuperspace  except on its boundary. In particular, it
does not imply that the wave function is regular when the
three--volume is vanishingly small. This fact leads us to study
the boundary $\partial\Omega$ of the minisuperspace  $\Omega$.
According to Vilenkin \cite{vi88}, we shall define it as the
set of all configurations which are singular in a general
sense, i.e. such that make the metric $f_{\mu\nu}/N$ singular.
On the one hand, $\partial\Omega$ may contain configurations
whose singularities are due to the slicing procedure of a
regular Euclidean wormhole four--geometry but that are not
truly singular. This part of the boundary will be called
regular boundary. In particular, the configurations associated
with vanishing three--volume belong to the regular boundary. On
the other hand, the remaining part of the boundary, the so
called singular boundary, will consist of real singularities in
minisuperspace which will not be due to  the spacetime
foliation. It should be mentioned that there may exist singular
three--geometries which cannot be embedded in any regular
four--geometry. However, they will not appear when a slicing of
a regular four--geometry is performed, and therefore they will
not be relevant to the quantization procedure.

The square integrability boundary condition  concerns to the
singular boundary, but it says nothing about the regular
boundary. In this respect, two questions arise concerning the
regular boundary:  $(i)$ Can we properly impose wormhole
boundary conditions  on it? and $(ii)$ Is it a real boundary of
minisuperspace? The answer to both questions is negative. The
Wheeler--DeWitt operator is hyperbolic and, for this kind of
operators, the boundary value problem is well posed only if
boundary conditions are imposed on characteristic surfaces of
the configuration space \cite{pu55}, that is on surfaces
$u(q)=0$ such that their normal vectors are null
\[
f^{\mu\nu}\nabla_\mu u\nabla_\nu u=0 \ .
\]
We will  see that the regular boundary is not a characteristic
surface in general (this will be the case, for instance, in the
minisuperspace  model  with a conformal scalar field  of
Sec.~\ref{sec:mss}). This question is closely related to
whether the Wheeler--DeWitt operator is self--adjoint or not.
In fact, the Wheeler--DeWitt operator is essentially
self--adjoint and an appropriate choice of boundary conditions
will determine its self--adjoint extension, provided that the
boundary value problem is well posed.  With respect to the
second question, we have seen that the regular boundary is made
of `coordinate singularities' rather than true singularities.
It plays a similar role to the Schwarzschild horizon in black
hole physics. It may be preventing us from gaining  access to
other regions of the configuration space.

For the reasons explained above, one is naturally led to
consider the maximal analytic extension \cite{he73} $\bar
\Omega$ of minisuperspace  as  the configuration space in which
the quantization procedure can be properly accomplished. We can
now state the wormhole boundary conditions  as saying that the
wormhole wave functions must be square integrable in the
maximal analytic extension of minisuperspace. In other words,
the Hilbert space of wormholes ${\cal W}^{\rm o}$ consists of
all normalizable solutions of the Wheeler--DeWitt equation
\[
{\cal W}^{\rm o}=\left\{\Psi(\bar q)\in L_{\bar
f}^2(\bar\Omega),\hspace{.5cm}
\hat{\bar H}\Psi(\bar q) =0\right\} \ ,
\]
where the inner product is given by the natural measure in the
extended minisuperspace  $d\bar q\sqrt{\bar f}$, the overbar
denotes quantities in the extended minisuperspace  and $\bar f$
is the determinant of the extended metric.

This boundary condition  gives full meaning to the wormhole
boundary conditions  formulated previously. Indeed, the
boundary of the maximal extension of minisuperspace  is truly
singular and therefore the wormhole wave function will vanish
there by virtue of square integrability. Besides, the regular
boundary of minisuperspace  (and in particular the
configurations that represent vanishing three--geometries) is
in the interior of the extended minisuperspace  and thus no
additional boundary conditions  are required there. Regularity
of the  wave function in these configurations is automatically
guaranteed. As mentioned above, this boundary condition
manifestly shows the absence of a spacetime character in
minisuperspace  and, in particular, the non--existence  of a
time variable  despite the Lorentzian signature of the metric
in minisuperspace. In this sense, the Wheeler--DeWitt operator
is like the Hamiltonian of a Schr\"odinger equation rather than
a Klein--Gordon operator.

In the examples that we will consider, the Wheeler--DeWitt
operator, $\hat{\bar H}$, defined on the extended
minisuperspace, is self--adjoint in $L_{\bar f}^2(\bar\Omega)$,
but this may not be the general case. The boundary of the
extended minisuperspace consists of true singularities of the
minisuperspace metric. These singular configurations  are
associated to true singularities of the four--geometry and, in
particular, the boundary of $\bar\Omega$ will contain the
configurations for which the three--geometry is asymptotically
large. Therefore, the wave function must vanish in that
boundary. In general, extended minisuperspace variables will
run along the whole real line and this boundary will be at
infinity. In this case, the Wheeler--DeWitt operator will be
self--adjoint in  $L_{\bar f}^2(\bar\Omega)$. If any of the
characteristic surfaces that define the boundary of
minisuperspace is not located at infinity, an additional
boundary condition in that surface will be necessary, so that
$\hat{\bar H}$ be self--adjoint. The square integrability of
the wave function  will not be sufficient and we will have to
explicitly impose the condition that it vanishes in such a
surface. In the minisuperspace models that we have considered,
the square integrability condition is sufficient.


\section{Hilbert space structure}
\label{sec:hs}

In order to acquire a deeper understanding of this Hilbert space
structure, let us suspend the Hamiltonian constraint for a
while. Let us define an inner product in the space $\cal W$ of
functions in $\bar\Omega$ which satisfy wormhole boundary
conditions (note that these functions need not be anihilated by
the Wheeler--DeWitt operator $\hat{\bar H})$ as
\[
(\Psi_2,\Psi_1)=\int_{\bar\Omega}\!d\bar q\sqrt{\bar f}\
\Psi_2^*(\bar q)\Psi_1(\bar q) \ ,
\]
so that ${\cal W}=L_{\bar f}^2(\bar\Omega)$. The Wheeler--DeWitt
operator is self--adjoint in this inner product as already
discussed. The boundary conditions in the characteristic
surfaces located at a finite distance cancel the surface terms
which appear when an integration by parts is performed in the
difference
\[
(\Psi_2,\hat{\bar H}\Psi_1)-(\hat{\bar H}\Psi_2,\Psi_1) \ .
\]
The spectrum of the Wheeler--DeWitt operator, $\sigma(\hat{\bar
H})$, will be real and will consist of an essential part and a
discrete one. Let $\Psi_{\omega,\xi}(\bar q)$ be its
eigenfunctions,
\[
\hat{\bar H}\Psi_{\omega,\xi}(\bar
q)=\omega\Psi_{\omega,\xi}(\bar q) \ ,
\]
where $\omega\in\sigma(\hat{\bar H})$ and the index $\xi$
distinguishes between different elements of an orthonormal basis
of the subspace ${\cal W}^\omega$ of eigenfunctions associated
to the eigenvalue $\omega$, provided that the multiplicity of
$\omega$ is different from unity. Then the set
$\left\{\Psi_{\omega,\xi}(\bar q)\right\}$  forms an orthonormal
basis of $\cal W$, i.e. its elements satisfy the conditions
\[
(\Psi_{\omega,\xi},\Psi_{\omega\prime,\xi\prime})=
\delta(\omega-\omega\prime)\delta(\xi-\xi\prime) \ ,
\]
\[
\int d\omega d\xi\  \Psi_{\omega,\xi}^*(\bar
q)\Psi_{\omega,\xi}(\bar  q\prime)=\frac{1}{\bar f}\ \delta(\bar
q-\bar q\prime) \ ,
\]
where the first relation expresses the orthonormal character of
the eigenfunctions and the second is the spectral decomposition
of the identity in terms of eigenfunctions of the
Wheeler--DeWitt operator. For the sake of simplicity in the
notation, $d\omega d\xi$ represents the measure in
$\sigma(\hat{\bar H})$ provided by the spectral theorem for
unbounded operators \cite{rs72}. In particular, if the spectrum
of $\hat{\bar H}$ is discrete, we will have to substitute
$d\omega$ by $d\omega\delta(\omega-\omega_n)$, being $\omega_n$
the discrete eigenvalues; that is, the integral will be
transformed into a discrete sum.

By means of the isomorphism
\begin{eqnarray}
\sigma:{\cal W}&\rightarrow&{\cal H}\nonumber\\
\Psi&\mapsto&\left|\psi\right\rangle \ ,\nonumber
\end{eqnarray}
we can assign to each basis element a state
$\left|\omega,\xi\right\rangle=\sigma\Psi_{\omega,\xi}(\bar q)$,
so that ${\cal H}$ is the set of all states
$\left|\psi\right\rangle$ which are linear combinations of the
basis elements
\[
\left|\psi\right\rangle=\int d\omega d\xi\
\psi(\omega,\xi)\left|\omega,\xi\right\rangle \ ,
\]
in which the coefficients are square integrable, i.e.
\[
\int d\omega d\xi\  |\psi(\omega, \xi)|^2=1 \ .
\]
 The space ${\cal H}^{\rm o}$, of states that satisfy the
Wheeler--DeWitt equation, is a subspace of ${\cal H}$ associated
to the eigenvalue $\omega=0$ of the operator $\hat{\bar H}$. If
the zero eigenvalue does not belong to the discrete spectrum of
the Wheeler--DeWitt operator, ${\cal H}^{\rm o}$  will not be a
true subspace of ${\cal H}$ and therefore it will not represent
the Hilbert space of wormhole wavefunctions. Indeed, if
$0\in\sigma_{\rm ess}(\hat{\bar H})$, then no state
$\left|0,\psi\right\rangle\in{\cal H}^{\rm o}$ will be
normalizable, since
$\left\langle0\psi|0,\psi\right\rangle=\delta(\omega=0)$.  In
all minisuperspace models that we have studied and those which
have appeared in the literature (see, for example, Refs.
\cite{ha88,hp90,cg91,do90,ly89,dl91,zh92}) the Wheeler--DeWitt
operator has a discrete spectrum $\sigma_{\rm disc}(\hat{\bar
H})$ so that $0\in\sigma_{\rm disc}(\hat{\bar H})$.  It should
be noted that the operators $\hat{\bar q}^\mu$, which represent
the three--geometry and the matter fields, are self--adjoint in
${\cal H}$. The spectrum of $\hat{ \bar q}^\mu$ is continuous
and therefore the eigenstates $\left|\bar q\right\rangle$ do not
belong, strictly speaking, to ${\cal H}$ but they form a
continuous basis of ${\cal H}$. However, since $\hat{\bar
H}\left|\bar q\right\rangle\neq 0$, the vectors $\left|\bar
q\right\rangle$ do not belong to ${\cal H}^{\rm o}$,  not even
in the sense of states associated to continuous eigenvalues.

\subsection{The Schr\"odinger equation}

In quantum cosmology, the suspension of the Wheeler--DeWitt
equation, the constraint which guarantees the invariance under
time reparametrizations, gives rise to an Schr\"odinger
equation. The wave function can be written  as a path integral
between two three--surfaces: in one of them we define the
arguments of the wave function while in the other we impose
boundary conditions.  This integral contains a sum over all
possible lapse functions.  The invariance of the action under
reparametrizations which do not affect to these three--surfaces
allow us to divide the whole set of lapse functions in
equivalence classes. It is necessary to introduce a condition
that fixes the gauge in this sum, so that equal contributions be
counted only once. If we have two three-surfaces, this condition
can be written in the form \cite{te83,ha88a}
\[
\dot N=0 \ .
\]
Any other history   $N(\tau)$ can be transformed into one which
satisfies the gauge condition. The final form of the wave
function is then
\[
\Psi^{\rm o}(\bar q)=\int_\gamma\!dN\Psi(\bar q,N) \ ,
\]
where the functional integral over all histories  $N(\tau)$ has
been substituted, due to the gauge fixing condition, by an
ordinary integral along a contour $\gamma$ in the complex
$N$--plane for which the integral is convergent and
\[
\Psi(\bar q,N)=\int{\cal D} \bar q{\cal D} \bar p\exp\left\{
-\int_0^N\!d\tau(\bar p_\mu\dot{\bar q }^\mu-\bar H)\right\} \ .
\]
The function  $\Psi(\bar q,N)$ satisfies the Euclidean
Schr\"odinger equation
\[
\hat{\bar H}\Psi(\bar q,N)=-\partial_N\Psi(\bar q,N)
\]
and therefore  $\Psi^{\rm o}(\bar q)$ verifies the
Wheeler--DeWitt equation $\hat{\bar H}\Psi^{\rm o}(\bar q)=0$,
provided that the contour $\gamma$ is invariant under
reparametrizations as when, for instance, it is closed or
infinite \cite{hh91}.

When the boundary conditions are of the wormhole type, one of
the surfaces is taken to infinity. Then the gauge fixing
condition is stronger \cite{ga91}:
\[
N=1 \ .
\]
Indeed, this condition determines the existence of one single
equivalence class under gauge transformations. Then the wormhole
wave function is
\[
\Psi^{\rm o}(\bar q)=\lim_{|T|\rightarrow\infty}\Psi(\bar q,T) \ ,
\]
where $T$ is the coordinate time distance between both surfaces
and
\[
\Psi(\bar q,T)=\int{\cal D} \bar q{\cal D}\bar
p\exp\left\{-\int_0^T\!d\tau
(\bar p_\mu\dot{ \bar q}^\mu-\bar H)+\mbox{s.t.}\right\} \ .
\]
The surface terms which appear in this expression are due to the
wormhole boundary conditions and were discussed in detail in
Ref. \cite{ga91}. As happens in quantum cosmology, $\Psi(\bar
q,T)$ satisfies the Euclidean Schr\"odinger equation.  However,
the Wheeler--DeWitt equation is not obtained as a consequence of
the integration over $T$, but by taking the limit
$T\rightarrow\infty$, i.e. driving the surface in which boundary
conditions are defined to its original place: the asymptotic
region. Indeed, the function  $\Psi(\bar q,T)$  can be written
as a linear combination of eigenfunctions of $\hat{\bar H}$
\[
\Psi(\bar q,T)=\int d\omega d\xi\ \psi(\omega,\xi) e^{-\omega
T}\Psi_{\omega,\xi}(\bar q) \ .
\]
If  $\psi(\omega,\xi)\neq \psi(\xi)\delta(\omega)$, then terms
with $\omega>0$ will not contribute in the limit
$T\rightarrow\infty$ but terms  with $\omega<0$ will give an
infinite contribution (the situation is reversed in the limit
$T\rightarrow-\infty$). Thus for the wormhole wave function
$\Psi^{\rm o}$ to be well defined, the only contribution that
can survive is $\omega=0$, that is, the Wheeler--DeWitt operator
must anihilate the wave function $\Psi^{\rm o}$, $\hat{\bar
H}\Psi^{\rm o}(\bar q)=0$.  This heuristic argument makes use of
one of the main features of quantum gravity: the unboundedness
from below of the gravitational action. In the Hamiltonian
formulation, this feature manifests itself  in the fact that the
Wheeler--DeWitt operator is hyperbolic and therefore  it is not
unbounded from below, so that it admits arbitarily negative
eigenvalues.

Both in quantum cosmology and wormholes, the Wheeler--DeWitt
equation is satisfied due to the reparametrization invariance
which enforces a sum over all lapse functions. However, due to
the different nature of the boundary conditions in both
situations, the way of performing this sum is different. In
quantum cosmology, we have to sum over all possible time
separations.  In turn, in wormhole physics, there is only one
time separation, whose value is infinity, between the surface in
which the arguments of the wave function are defined and the
surface in which we impose the boundary conditions \cite{ga91}.

It is not easy to find a physical  interpretation for this
Schr\"odinger equation. Formally, it determines the evolution of
the wave function in a time which is not observable, since all
observable physical quantities are expressed as sums over all
possible times. However, the spectral theory of the
Wheeler--DeWitt operator gives information, not only about
wormholes, but also about the structure of quantum cosmology
itself; that is, considering off shell configurations and thus
suspending the Hamiltonian constraint, permits the calculation
of Green's functions of the Wheeler--DeWitt operator. Given two
configurations  $\bar q_1$ and $\bar q_2$ of extended
minisuperspace, the transition amplitude between them both is
defined by the path integral
\[
G(\bar q_1,\bar q_2)=\int_\gamma\! dN\int_{{\cal C}_{12}}\!{\cal
D} \bar q\ e^{-I[\bar q,N]} \ ,
\]
where ${\cal C}_{12}$ is the set of histories $\bar q(\tau)$
such that $\bar q(0)=\bar q_1$ and $\bar q(1)=\bar q_2$ and
$\gamma$ is an integration  contour in the complex $N$--plane.
If $\gamma$ is infinite or closed, $G(\bar q_1,\bar q_2)$ will
be a solution of the Wheeler--DeWitt equation but, if $\gamma$
is semiinifinite, then $G(\bar q_1,\bar q_2)$ will be a Green's
function of the Wheeler--DeWitt operator, that is,
\[
\hat{\bar H}(\bar q_1)G(\bar q_1,\bar q_2)=\frac{1}{\bar f}\
\delta(\bar q_1-\bar q_2) \ .
\]
We can write, at least formally, the Green's function of the
Wheeler--DeWitt operator as a sum over  eigenstates of
$\hat{\bar H}$
\[
G(\bar q_1,\bar q_2)=\int d\omega d\xi\  \Psi_{\omega,\xi}(\bar
q_1)\frac{1}{\omega}  \Psi_{\omega,\xi}^*(\bar q_2) \ .
\]
Indeed, the expression
\[
g(\bar q_1,\bar q_2;N)=\int_{{\cal C}_{12}}\!{\cal D} \bar q\
e^{-I[\bar  q,N]}
\] is a propagator in ordinary quantum mechanics and therefore
it may be writen in terms of the eigenfunctions of the
Hamiltonian as
\[
g(\bar q_1,\bar q_2;N)=\int d\omega d\xi\ \Psi_{\omega,\xi}(\bar
q_1)e^{-\omega N}\Psi_{\omega,\xi}^*(\bar q_2) \ .
\]
The integration over $N$ along a semiinfinite contour  $\gamma$
in which the integral converges gives the expected result.


\subsection{The effective interaction}

In the study of the efects that wormholes exert on the low
energy physics, it is necessary to analyze the  matter field
propagator
\[
\left\langle0\right|\Phi(x_1)\cdots\Phi(x_r)\Phi(y_1)
\cdots\Phi(y_s)\left|0\right\rangle
\]
between two asymptotically Euclidean regions \cite{ha88}. The
points  $x_1\ldots x_r$ are located in one of these regions and
the points  $y_1\ldots y_s$ in the  other. The state $\left|
0\right\rangle$ represents the ordinary scattering vacuum in
quantum field theory in flat spacetime. For the sake of
simplicity we will denote the product $\Phi(x_1)\cdots\Phi(x_r)$
simply by $\underline\Phi(x)$ and similarly for the fields at
the points $y_1\ldots y_s$. We can express this  propagator as
the  path integral over all four--metrics which can be
interpolated between two asymtotically Euclidean regions and
over matter fields that reach the vacuum configuration in both
regions
\[
\left\langle  0\right|\underline\Phi(x)\underline\Phi(y)
\left| 0\right\rangle=\int{\cal D}
g_{\alpha\beta}{\cal D}\Phi\
\underline\Phi(x)\underline\Phi(y)
e^{-I[ g_{\alpha\beta},\Phi]}\ .
\]
The action  $I[ g_{\alpha\beta},\Phi]$ must contain the
appropriate surface terms that take into account the
asymptotically Euclidean character of spacetime, as discussed in
Ref. \cite{ga91}.

In order to perform this integral, we must first eliminate all
those configurations which are related by gauge transformations
and that therefore leave the action unchanged. That is, we need
a condition that fixes the gauge, analogous to those discussed
above for the propagation between two three--geometries and for
the propagation between one three--geometry and an
asymptotically Euclidean region.  In this latter, the gauge
condition is much stronger than in the former. Indeed, when we
have two finite three--metrics the path integral over lapse
functions is reduced to an ordinary integral, while when one of
the surfaces corresponds to an asymptotically Euclidean region
the gauge condition reduces the path integral to a single term.
Now, both extreme configurations are asymptotically Euclidean.
There does not exist an appropriate  gauge fixing condition
since whatever may be the choice we make, there will be various
transformations that connect  any history with one satisfying
the gauge condition and therefore it is not possible to fix the
gauge completely. We are facing an ambiguity of the Gribov type
\cite{gr78}. Its cause may be the insistence on eliminating all
the gauge freedom with a single global condition over all the
spacetime \cite{ra90}.

A possible way of avoiding this ambiguity consists on
introducing a three--section $\Sigma$ that divides the spacetime
manifold into two disconected parts, each of them containing one
of the asymptotic regions. The spectral decomposition of the
identity in terms of the eigenstates of the Wheeler--DeWitt
operator over the surface  $\Sigma$,
\[
\openone=\int d\omega d\xi\left|\omega,
\xi\right\rangle\left\langle \omega,\xi\right| \ ,
\]
allows us to factorize the propagator
\[
\left\langle  0\right|\underline\Phi(x)\underline\Phi(y)
\left| 0\right\rangle=\int d\omega d\xi
\left\langle  0\right|\underline\Phi(x)
\left|\omega,\xi\right\rangle\left\langle \omega,\xi\right|
\underline\Phi(y)\left| 0\right\rangle \ ,
\]
where each of the factors is given by the path integral over
four--metrics and matter fields whose energy--momentum tensor
vanishes at infinity, anologous to that defining an on shell
wormhole wave function. Therefore as we have seen above  only
terms with  $\omega=0$ will give  finite non--vanishing
contribution, that is,
\[
\left\langle  0\right|\underline\Phi(x)\underline\Phi(y)
\left| 0\right\rangle=\int  d\xi
\left\langle  0\right|\underline\Phi(x)\left|0,\xi
\right\rangle\left\langle 0,\xi\right|\underline\Phi(y)
\left| 0\right\rangle \ ,
\]
where the states  $\left|0,\xi\right\rangle$ form a basis of the
subspace ${\cal H}^{\rm o}$ of wormholes  which satisfy the
Wheeler--DeWitt equation
\cite{ha88}. This corresponds to the idea that, in order to
study the effect of wormholes on the low energy fields, one just
has to  introduce a complete set of on shell wormhole   states
between both asymptotic regions. These states must be  solutions
of the Wheeler--DeWitt equation  since wormholes cannot carry
energy. It is also a consequence of the dilute wormhole
approximation. If the presence of other wormholes cannot be
ignored, the section $\Sigma$ will not divide the spacetime
manifold into two disconnected parts, being necessary the
density matrix formalism \cite{go91}, in which the spectrum of
the Wheeler--DeWitt operator plays a central role, in a similar
way  to what happened in the calculation of Green's functions.


\subsection{The lapse function}

The quantum formulation of wormholes depends on the choice of
the lapse function. Here we will analyze this dependence and see
to what extent this choice may affect our results. In the more
general context of quantum cosmology, the presence of this
ambiguity has already been pointed out by Hawking and Page
\cite{hp86}.

Let us write down the Friedmann--Robertson--Walker metric in the
form
\[
ds^2=\frac{2\mbox{\small G}}{3\pi}\left( N^2{\cal
N}^2(a)d\tau^2+a^2d\Omega_3^2\right) \ ,
\]
where ${\cal N}(a)$ is a function of $a$ which determines the
gauge and  $N$ is the lapse function in this gauge which,
without loss of generality, will be regarded as constant. For
the sake of simplicity, we will consider only functions of the
form
\[
{\cal N}(a)=a^z \ ,
\hspace{.5cm}\hspace{.5cm}
z\in I\!\! R \ .
\]
In particular, $z=0$ corresponds to the proper time gauge and
$z=1$ to the conformal gauge. The relation between the proper
time $t$ and the coordinate time $\tau$  is given by the
expression $dt=N{\cal N}\left( a\right) d\tau$.

Let us start studying which is the effect of the choice of
${\cal N}(a)$ on the regularity condition for three--geometries
of vanishing volume. The wave function must be regular at $a=0$,
expressing the fact that the spacetime manifold is not singular.
Then such manifold will admit a tangent plane at the point for
which $a=0$, that is, $a\sim t$ when the proper time $t$ goes to
zero. In terms of the coordinate time $\tau$, this behaviour
will be determined by the equation $\dot a(\tau)\sim N{\cal
N}\left( a\right)$.  This means that
\[
\begin{array}{rcll}
a(\tau)^{1-z} &\sim& (1-z)N\tau&
\hspace{.5cm}\hspace{.5cm}(z\neq1)\\
a(\tau)&\sim&\exp(N\tau)&\hspace{.5cm}\hspace{.5cm}(z=1) \ .
\end{array}
\]
Consequently, the point $a=0$ is labelled by a infinite negative
coordinate time if $z>1$ and finite $z<1$. The case $z=1$ is the
limit in which  $a=0$ corresponds to an infinite negative
coordinate time but with an exponential behaviour and,
therefore, faster than the inverse of any power. This has a
simple interpretation. The function ${\cal N}(a)$ defines the
density of the foliation  in the ($3+1$) formalism, that is, the
number of leaves per unit proper time. Indeed, this number is
given by the ratio $dt/d\tau$ between the interleave distances
in both foliations, which is precisely the function ${\cal
N}(a)$. The foliation associated to the conformal gauge $z=1$ is
a limiting case, as we have seen. The foliations associated to
values of $z$ greater than unity are less dense than the
conformal one and those associated to $z$ smaller than one are
more dense.

{}From the classical point of view, this kind of descriptions are
certainly valid. However, in the canonical quantum theory some
of these functions ${\cal N}(a)$ are inadmissible since they
give rise to pathologies without any physical meaning.  Indeed,
the requirement that the Wheeler--DeWitt operator be
self--adjoint selects the scalar product and, therefore, the
measure in minisuperspace. A consistent choice of the operator
ordering and the integration measure has been discussed in the
precceding sections: the volume element as integration measure
and the  operator ordering in $\hat H$ such that the kinetic
term has the form of the natural Laplacian in such measure.  The
measure will be $da a^2{\cal N}(a)^{-1}$, or more explicitly
$ada a^{1-z}$. Taking into account that  the three--metric is $
q_{ik}=a^2\Omega_{ik}$, we can interpret $ada$ as the analog of
the measure in the space of three--metrics ${\cal D} q_{ik}$. We
can see then that if $z>1$, small volume metrics will be
weighted by an extremely large  factor, due to the low density
of the foliation in that region, thus giving them and importance
which they do not deserve since zero volume configurations do
not represent special points of the spacetime manifold  at all.

This discussion is based on the requirement that the
Wheeler--DeWitt operator be self--adjoint. In ordinary quantum
mechanics  this requirement guarantees that the eigenvalues of
the Hamiltonian are real since they are possible results of
measurements.  In our case, wormhole wave functions are
anihilated by the Wheeler--DeWitt operator  and it may seem
somewhat unnecessary the requirement that this operator be
self--adjoint. However a simple argument based on the boundary
conditions will convince us that this is not the case. Let
$\Psi_1^{\rm o }$ and $\Psi_2^{\rm o }$ be two arbitrary wave
functions which are solutions of the Wheeler--DeWitt equation.
Then,
\begin{eqnarray}
&&\left(\Psi_2^{\rm o },\hat{\bar H}\Psi_1^{\rm o
}\right)-\left( \hat{\bar H}\Psi_2^{\rm o },\Psi_1^{\rm o
}\right) \nonumber\\
&&=\int_{\partial\bar\Omega}\!d\bar\sigma^\mu\left(\Psi_2^{\rm
o*}\bar\nabla_\mu\Psi_1^{\rm o}- \Psi_1^{\rm
o}\bar\nabla_\mu\Psi_2^{\rm o*}\right) \ . \nonumber
\end{eqnarray}
For $\hat H$ to  be self--adjoint, both surface terms must be
equal. This is in fact the case since boundary conditions define
the behaviour of the wave functions in the boundary of
minisuperspace, $\partial\bar\Omega$, and this behaviour is the
same for them all. Thus well defined boundary conditions
guarantee the self--adjointness of the Wheeler--DeWitt operator.

We can state that the gauge choices ${\cal N} (a)$ that vanish
too fast when the three--geometry degenerates will not produce a
well defined canonical quantum wormhole theory since, in these
cases, the minisuperspace measure will be infinite at the
configurations associated to vanishing volume three--geometries,
in contradiction with the meaning of the wormhole boundary
conditions. From the more general point of view of the spectrum
of the Wheeler--DeWitt operator, these gauge choices  imply
$\sigma_{\rm disc} (\hat{\bar H})=\emptyset$ and therefore that
the zero eigenvalue is in the essential spectrum, $0\in
\sigma_{\rm ess}(\hat{\bar H})$, associated to non--normalizable
wave functions. We will say that these gauge fixing conditions
are inadmissible from the canonical quantum point of view.

It should be stressed that different choices of the function
${\cal N}(a)$, even though they are admissible, will give rise
to different canonical theories, not because the structure of
the Wheeler--DeWitt operator or the path integral which define
the wave function will change but because of the construction of
the Hilbert space in which the wormhole wave functions live
since its scalar product depends on this choice. That is,
although the local laws in minisuperspace do not depend on the
choice of ${\cal N} (a)$, the formulation of the global laws, of
boundary conditions, and therefore the Hilbert space structure
do.  However, qualitative results such as the absence of giant
wormholes, which is discussed in the next section, survive to
this ambiguity.

On the other hand, effective interactions are defined in terms
of path integrals which contain sums over all posible lapse
functions. Choosing a function  ${\cal N} (a)$ corresponds to a
partial gauge fixation in these integrals, which determine the
maximal analytic extension of minisuperspace. Since the gauge
fixation is carried out in such a way that the path integral is
independent of the gauge fixing condition, it will be possible
to perform the canonical quantization in any of these gauges,
provided they are admissible. The final result for the effective
interaction will be independent of this choice.


\section {Minisuperspace models}
\label{sec:mss}

In this  section, we will illustrate some aspects of the
wormhole Hilbert space structure in some particular
minisuperspace  models.  Let us consider a
Friedmann--Robertson--Walker spacetime whose metric, in the
conformal gauge (${\cal N}(a)=a$, $N=1$), can be written as
\[
ds^2=\frac{2\mbox{\small G}}{3\pi} a^2(\tau)\left(
d\tau^2+d\Omega^2_3\right).
\]
As the matter content, we will consider a minimally or
conformally coupled massless homogeneous scalar field.


\subsection{Minimal coupling}

The Euclidean action, when the scalar field is minimally coupled
to a Friedmann--Robertson--Walker spacetime, has the form
\[
I=\int_0^\infty\!\!d\tau \left\{-\frac{1}{2}\dot a^2-\frac{1}{2}
a^2+\frac{1}{2} a^2\dot\phi^2\right\} \ ,
\]
where $\phi(\tau)$ is the minimal scalar field.  Therefore, the
metric in minisuperspace  is
\[
d{\cal S}^2= -da^2+a^2d\phi^2
\]
and the boundary of minisuperspace  will consist of
\[
\partial\Omega=\left\{ (a,\phi),\ \ a=\infty \ \ \mbox{or}\ \
\phi=\infty \ \ \mbox{or}\ \ a=0\right\} \ .
\]
Although $f^{\mu\nu}$ is not well defined at $a=0$, the singular
character of this configuration  is due to the foliation of
spacetime. Then it may be expected that it belongs to the
regular boundary of minisuperspace. On the other hand,
$a=\infty$ induces an infinite volume, both in minisuperspace
and spacetime. This suggests that it is truly singular. This is
in fact the case. The change of variables
\[
x=a\cosh\phi \ ,
\hspace{.5cm}\hspace{.5cm}
y=a\sinh\phi \ ,
\]
transforms the minisuperspace  metric into the Minkowski metric
\[
d{\cal S}^2=-dx^2+dy^2 \ ,
\]
although only region I in Fig.~\ref{fig:min} is represented
since in these variables the minisuperspace  is
\[
\Omega=\left\{ (x,y),\ \ x>|y|\right\} \ .
\]
Part of the boundary of minisuperspace  $\partial\Omega$, that
defined by the points such that $x=|y|$, consists of
configurations which are not singular. Therefore, it seems
natural to extend the range of the variables $x$ and $y$ beyond
the boundary of minisuperspace. The maximal analytic extension
$\bar\Omega$ of the minisuperspace  $\Omega$  is then given by
the whole Minkowski plane (Fig.~\ref{fig:min}) and constitutes
the natural basic space for the quantization of this system.

The wormhole boundary conditions which have been proposed in the
previous sections  require that the wave function be square
integrable, i.e. that  $\Psi(x,y)\in L^2( I\!\! R^2)$. A direct
consequence of these conditions is that the operators $\hat x,\
\hat y,\ i\hat p_x$ and $i\hat p_y$ (the imaginary unit appears
because we are dealing with Euclidean momenta) are self--adjoint
in their respective domains and besides form Schr\"odinger pairs
\cite{rs72,gp89}.   Analogously, the Wheeler--DeWitt operator
\[
\hat{\bar H}=-\partial_x^2+\partial_y^2+x^2-y^2
\]
is self--adjoint in $L^2( I\!\! R^2)$. On the other hand, the
restriction to the original minisuperspace  causes the indices
$d_\pm(\hat p_a)= \dim\left(\ker\left\{ \hat p_a\pm
1\right\}\right)$ of the operator $i\hat p_a$  not to coincide
and, therefore, the operator $i\hat p_a$ will not be essentially
self--adjoint in $C^\infty( I\!\! R_+)$  \cite{rs72,gp89}. In
spite of this fact, the operator
$-a^{-1}\partial_aa\partial_a$  is essentially self--adjoint and
consequently admits self--adjoint extensions. By means of
suitable boundary conditions at $a=0$, one can choose the
adecuate extension \cite{gp89}. The election of this
self--adjoint extension gives a precise meaning to the
regularity condition proposed by Hawking and Page  \cite{hp90}
in this model since the self--adjointness of the Wheeler--DeWitt
operator
\[
\hat
H=-\frac{1}{a}\partial_aa\partial_a+
\frac{1}{a^2}\partial_\phi^2+a^2
\]
is determined in this case by that of the operator
$-a^{-1}\partial_aa\partial_a$.  This one is a Sturm--Liouville
operator  which is singular at $a=0$. The most general boundary
conditions that may be imposed at $a=0$ so that it be
self--adjoint are of the form
\[
\lim_{a\rightarrow 0}\left\{\alpha\Psi-(\alpha\log
a-1)a\partial_a\Psi\right\}=0 \ ,
\]
where $\alpha$  is an arbitrary parameter that determines the
self--adjoint extension. To choose the value of the parameter
$\alpha$,  we shall look at the extended minisuperspace
$\bar\Omega$. A basis of wave functions which are square
integrable is given in Ref. \cite{hp90}
\[
\Psi_n^{\rm o}(x,y)=\varphi_n(x)\varphi_n(y) \ ,
\hspace{.5cm}\hspace{.5cm}
n=0,1,2\ldots
\]
where  $\varphi_n(x)=\left( n!2^n\right)^{-1/2}H_n(x)e^{-x^2/2}$
are the eigenfunctions of the Hamiltonian of the harmonic
oscillator $-\partial_x^2+x^2$.  If $\alpha\neq0$, only the
functions $\Psi_n^{\rm o}$ for which  $n$ is even will  satisfy
the boundary condition at $a=0$. In this case the domain of the
self--adjoint extension of the Wheeler--DeWitt operator will be
too small. Indeed, not only have we excluded part of the
functions $\Psi_n^{\rm o}$, but we have not included the wave
functions  $\Psi_{\phi_0}^{\rm o}$  which were obtained by
direct path integration in Ref. \cite{ga91} either.  The
boundary condition defined by $\alpha=0$ is
\[
a\partial_a\Psi(a,\phi)\vert_{a=0}=0 \ .
\]
Every wave function which belongs to  $L_{\bar f}^2(\bar\Omega)$
satisfies this condition as shown below. Therefore, in the
restricted minisuperspace   $\Omega$, the boundary conditions
that the wave function be regular when the three--geometry
degenerates and that it be damped for asymptotically large
three--geometries can be formulated in the single boundary
condition
\[
\Psi^{\rm o}(a,\phi)\in L_f^2(\Omega) \ .
\]
On the other hand, the formulation in the extended
minisuperspace  naturally contains to this condition. Indeed,
the zero volume three--geometry is represented in $\bar\Omega=
I\!\! R^2$ by the points $|x|=|y|$, i.e. by the null cone of the
origin in the Minkowski metric. The normal derivatives to the
null cone of the basis wave functions $\left\{\Psi_n^{\rm
o}(x,y)\right\}$ vanish. However, this is a condition which has
naturally appeared when requiring that the wave function be
square integrable in the maximal analytic extension of
minisuperspace  and which has not been imposed {\it a priori}.
In this model, the formulation of the boundary conditions in the
extended minisuperspace  has allowed us to clarify the meaning
of the regularity condition of the wave function in
configurations of zero volume although the quantization process
could also have been carried out consistently in the restricted
minisuperspace. However, in other models such as the one
presented below, it is necessary to consider the maximal
analytic extension of minisuperspace   in order to formulate the
quantum theory in a consistent way, as described in the previous
sections.

Apart from the discrete basis given by the wave functions
$\Psi_n^{\rm o}(x,y)$, there exists another continuous basis
$\left\{\Psi_k^{\rm o}(x,y)\right\}$  whose elements, written as
a linear combinations of the discrete ones, have the form
\[
\Psi_k^{\rm o}(x,y)=\sum_{n=0}^\infty \psi_n(k)\Psi_n^{\rm
o}(x,y) \ ,
\]
where the coefficients $\psi_n(k)$ are
\[
\psi_n(k)=\sqrt{\cosh\pi k/2}\int d\eta
\frac{\sinh^n\eta}{\cosh^{n+1}\eta} e^{-ik\eta} \ .
\]
Since they form a continuous basis, they are not square
integrable, but satisfy the closure relation
\[
\int\! dxdy {\Psi_k^{\rm o}}^*(x,y)\Psi_{k\prime}^{\rm o}(x,y)=
\sum_n\psi_n^*(k)\psi_n(k\prime)=\delta(k-k\prime) \ .
\]
The explicit expression of these wave functions is
\[
\Psi_k^{\rm o}(x,y)=\sqrt{\cosh\pi
k/2}K_{i\frac{k}{2}}\left(|x^2-y^2|/2\right) e^{-ik T(x,y)} \ ,
\]
where $K_{ik}(x)$ is the modified Bessel function of order $ik$
\cite{as65} and
\[
 T(x,y)=\left\{\begin{array}{ll}\tanh^{-1}(x/y)&\mbox{if
$|x|<|y|$}\\
\tanh^{-1}(y/x)&\mbox{if $|x|>|y|$}\end{array}\right. \ .
\]
The restriction to minisuperspace $\Omega$ acquires the simpler
form, as a function of the variables $a$ and $\phi$,
\[
\Psi_k^{\rm o}(a,\phi)=\sqrt{\cosh\pi k/2}\ K_{i\frac{k}{2}}\left(
a^2/2\right) e^{-ik\phi} \ .
\]
The behaviour  of this  function close to the boundaries is
\widetext
\[
\Psi_k^{\rm o}(a,\phi)\sim\frac{1}{\sqrt{k\tanh\pi k/2}}\left(
e^{i\theta(k)} e^{ik(\log a-\phi)} + e^{-i\theta(k)}e^{-ik(\log
a+\phi)}\right) \ ,
\]
\narrowtext
\noindent
when $a\rightarrow0$, where $\theta(k)$  is a real function of
$k$ whose form is irrelevant, and
\[
\Psi_k^{\rm o}(a,\phi)\sim\sqrt{\cosh\pi k/2}\
\frac{1}{a}e^{-a^2/2}e^{-ik\phi} \ ,
\]
when $a\rightarrow\infty$.  Therefore, they are exponentially
damped for asymptotically large  three--geometries. We can see
that the fact that they are not square integrable is due to its
non--regular behaviour in the origin, which is that of plane
waves in the variables $\phi,\ \log a$. However, they generate
the whole space of solutions of the Wheeler--DeWitt equation
which are square integrable. Indeed, let
\[
\Psi^{\rm o}(a,\phi)=\int dk\psi(k)\Psi_k^{\rm o}(a,\phi)
\]
be an arbitrary linear combination such that $\psi(k)\in L^2(
I\!\! R)$. Then,
\widetext
\[
a\partial_a\Psi^{\rm o}(a,\phi)\sim \int dk\ \psi(k)
\frac{k}{\sqrt{k\tanh\pi k/2}} \left( e^{i\theta(k)} e^{ik(\log
a-\phi)}  - e^{-i\theta(k)} e^{-ik(\log a+\phi)}\right)
\]
\narrowtext
\noindent
will be a function that, in the limit  $\log
a\rightarrow-\infty$ ($a\rightarrow0$), will vanish since
$\psi(k)$ is square integrable.

The value $a=\sqrt{|k|}$ corresponds to the wormhole throat
radius \cite{hp90}, since it separates the region in which the
wave function decreases exponentially from the region in which
it oscillates (see Fig.~\ref{fig:bes}).  The set of wormhole
wave functions which contain all the solutions which are square
integrable,
\[
{\cal W}^{\rm o}=\left\{\Psi^{\rm o}=\int dk\ \psi(k)\Psi_k^{\rm
o},
\hspace{.5cm}\int dk\ |\psi(k)|^2=1\right\} \ ,
\]
forms a Hilbert space which is isomorphic to $L^2( I\!\! R)$.
Note that this space is not  ${\cal W}=L_f^2(\Omega)$.  Indeed,
neither the functions  $\Psi_k^{\rm o}$ nor $\Psi_n^{\rm o}$
generate the whole set of square integrable functions in
$\Omega$ but only those which are eigenstates of the
Wheeler--DeWitt operator with eigenvalue zero.  As a
consequence,  it can be seen by direct calculation that they do
not satisfy the closure relations of $L_f^2(\Omega)$.

If, by means of the restriction of the  isomorphism $\sigma$
defined in Sec.~\ref{sec:hs} to ${\cal W}^{\rm o}$, we associate
the state $\left|0,k\right\rangle$ to the element $\Psi_k^{\rm
o}(x,y)$ of the basis of ${\cal W}^{\rm o}$, then the
coefficients of the expansion of
$\left|0,\psi\right\rangle=\sigma\Psi^{\rm o}(x,y)$ in terms of
$\left|0,k\right\rangle$ will be the same as those of $\Psi^{\rm
o}(x,y)$ in terms of $\Psi_k^{\rm o}(x,y)$. The scalar product
in  ${\cal H}^{\rm o}$ is induced by  $\sigma$ from the inner
product in  ${\cal W}^{\rm o}$, in such a form that
\[
\left\langle0,k\prime|0,k\right\rangle=\delta(k-k\prime).
\]
The identity operator in  ${\cal H}^{\rm o}$ can be decomposed
into a sum of projection operators on the basis elements,
\[
\openone_{{\cal H}^{\rm o}}=\int dk\left|0,k
\right\rangle\left\langle 0,k\right|,
\]
which induces the following closure relation in ${\cal W}^{\rm
o}$
\begin{eqnarray}
&&\openone_{{\cal W}^{\rm o}}\Psi^{\rm
o}(x_1,y_1)=\sigma^{-1}\openone_{{\cal H}^{\rm o}}\sigma\
\Psi^{\rm o}(x_1,y_1)\nonumber\\ &&=\int dk\ \Psi_k^{\rm
o}(x_1,y_1)\int dx_2dy_2\ {\Psi_k^{\rm o}}^*(x_2,y_2) \Psi^{\rm
o}(x_2,y_2). \nonumber
\end{eqnarray}
In contrast with usual procedures, it is crucial to mantain the
order of integration. Indeed, if the integration over $k$ is
first performed \cite{gr65}, an incorrect result will be
obtained.  This fact is closely related to the fact that there
does not exist any eigenstate of the operators  $\hat x,\ \hat
y$, or equivalently, $\hat a,\ \hat \phi$ in  ${\cal H}^{\rm
o}$. Indeed, although the operators $\hat x$ and $\hat  y$  are
self--adjoint in ${\cal W}=L^2(\bar\Omega)$, the subspace
${\cal W}^{\rm o}$ is not stable under the action of these
operators. It is easily seen that if the wave function
$\Psi^{\rm o}$ satisfies the Wheeler--DeWitt equation, then the
function  $(\hat x\Psi^{\rm o})(x,y)=x\Psi^{\rm o}(x,y)$ will
not satisfy it and likewise with $(\hat y\Psi^{\rm
o})(x,y)=y\Psi^{\rm o}(x,y)$.

Particularly interesting is the state
$\left|0,\phi_0\right\rangle$ defined by
\[
\left|0,\phi_0\right\rangle=\int dk\
\psi(\phi_0,k)\left|0,k\right\rangle \ ,
\]
where
\[
\psi(\phi_0,k)=\frac{1}{\sqrt{\cosh\pi k/2}} e^{ik\phi_0}
\]
or, in terms of the basis
$\left\{\left|0,n\right\rangle\right\}$, by
\[
\left|0,\phi_0\right\rangle=
\sum_n\psi_n(\phi_0)\left|0,n\right\rangle \ ,
\]
where
\[
\psi_n(\phi_0)=\frac{\sinh^n\phi_0}{\cosh^{n+1}\phi_0} \ .
\]
The value  $\phi_0$ represents the scalar field value in the
asymptotic region as can be derived from the fact that the wave
function $\Psi_{\phi_0}^{\rm o}(x,y)= \sigma^{-1}\left|
0,\phi_0\right\rangle$ has the form
\[
\Psi_{\phi_0}^{\rm o}(x,y)=\exp\left\{
-\frac{1}{2}(x^2+y^2)\cosh2\phi_0+xy\sinh2\phi_0\right\}
\]
or, in terms of the variables  $a$ and $\phi$,
\[
\Psi_{\phi_0}^{\rm o}(a,\phi)=\exp\left\{-\frac{1}{2}
a^2\cosh2(\phi-\phi_0)\right\} \ .
\]
This wave function has already been obtained as the path
integral over all asymptotically Euclidean metrics and over all
matter fields which reach the value  $\phi_0$ in the asymptotic
region \cite{ga91}.

We can introduce another continuous,  orthonormal basis
conjugate to $\left\{\left|0,k\right\rangle\right\}$, whose
elements are
\[
\left|0,\chi\right\rangle=
\int dk\ e^{ik\chi}\left|0,k\right\rangle
\]
which do not belong strictly to the space ${\cal H}^{\rm o}$. We
have seen that $\Psi_k^{\rm o}\in\!\!\!\!\!/\ {\cal W}^{\rm o}$
because it oscillates an infinite number of times at  $a=0$. The
functions  $\Psi_{\chi}^{\rm o}(a,\phi)$ are perfectly regular
at $a=0$, but oscillates for large values of $a$ and $\phi$. It
can be seen that
\[
\Psi_{\chi}^{\rm o}(a,\phi)=\cos\left(\frac{1}{2}
a^2\sinh2(\phi-\chi)\right)+ \mbox{terms}\in L_f^2(\Omega) \ .
\]
Although the operator  $\hat k$, whose eigenstates are
$\left|0,k\right\rangle$, represents both the flux through the
wormhole and the throat radius, there does  not seem to exist a
simple physical interpretation for the operator $\hat\chi$.
Nevertheless, both operators satisfy the canonical  conmutation
relation
\[
[\hat k,\hat \chi]=i\openone
\]
and besides form an Schr\"odinger  pair in the realization of
${\cal H}^{\rm o}$ as $L^2( I\!\! R)$. The discrete basis
$\left\{\left|0,n\right\rangle\right\}$ does not have a simple
interpretation either. Its elements are eigenstates,
corresponding to the same eigenvalue, of both the operators
$\hat H_x$ and  $\hat H_y$, which are formally equivalent to the
harmonic oscillator Hamiltonian and that, in the space ${\cal
W}^{\rm o}$, acquire the form $\hat H_x=-\partial_x^2+x^2,$
$\hat H_y=-\partial_y^2+y^2,$ so that
$\hat{\bar H}=\hat H_x-\hat H_y.$

The Hilbert space structure of the set of solutions of the
Wheeler--DeWitt equation with wormhole boundary conditions allow
us to introduce an interpretation for the wave function. Given a
wormhole in a state $\left|0,\psi\right\rangle$, the modulus
squared of the product with another state
$\left|0,\psi_0\right\rangle$,
$|\left\langle0,\psi|0,\psi_0\right\rangle|^2$, will give an
idea of the contribution of the state
$\left|0,\psi_0\right\rangle$ to the behaviour of the wormhole.
In particular,
$|\psi(k)|^2=|\left\langle0,k|0,\psi\right\rangle|^2$  indicates
in which proportion the behaviour corresponding to wormholes of
radius $\sqrt{|k|}$ appears. Since $|\psi(k)|^2\rightarrow0$
when $|k|\rightarrow\infty$, we can conclude that giant
wormholes (of arbitrary large radius) contribute  a neglible
proportion, and therefore will not dominate nor fill spacetime,
in agreement with the semiclassical results of Ref.
\cite{gs88}.

The fact previously stated that the eigenstates
$\left|x,y\right\rangle$ of the operators associated to  the
canonical variables $x$ and $y$ do not belong to ${\cal H}^{\rm
o}$, implies that this kind of interpretation is not applicable
to quantities such as $|\Psi^{\rm o}(x,y)|^2$.  This can be
easily understood since the configuration $(x,y)$ is defined in
a section of the four--geometry that defines the wormhole. The
wave function, however, defines global features of the whole
manifold, independent  of the section. It is nonsense,
therefore, to talk about ``a wormhole whose three--geometry is
defined by $x$ and $y$''.


\subsection{Conformal coupling}

Let  $\varphi(\tau)$ be a homogeneous and isotropic  scalar
field conformally coupled to a Friedmann--Robertson--Walker
spacetime whose metric has been written in conformal gauge. The
action of the system will be
\[
I=\int d\tau \left\{-\frac{1}{2}(1-\varphi^2)\dot
a^2-\frac{1}{2} a^2+
\frac{1}{2}a^2\dot\varphi^2 +a\varphi\dot
a\dot\varphi\right\} \ ,
\]
so that the line element in minisuperspace is
\[
d{\cal S}^2=-(1-\varphi^2)da^2+2a\varphi
dad\varphi+a^2d\varphi^2
\]
and, therefore, the metric and its inverse will be
\[
f_{\mu\nu}=\left(
\begin{array}{cc}
-(1-\varphi^2) & a\varphi\\ a\varphi & a^2
\end{array}
\right) \ ,
\]
and
\[
f^{\mu\nu}=\left(
\begin{array}{cc}
-1 & \varphi/a\\
\varphi/a & (1-\varphi^2)/a^2
\end{array}
\right) \ ,
\]
respectively. The values of $a$ for which the metric or its
inverse become infinite are  $a=\infty$ y $a=0$. The first value
corresponds to a three--geometry of infinite volume and
represents a true singularity.  The point $a=0$, which
corresponds to a three--geometry of zero volume, is not truly
singular; its singularity can be avoided by means of an analytic
extension. Finally, the values of the field $\varphi=\pm 1$ are
the limits in which the gravitational coupling changes its sign.
However, the metric in minisuperspace  is perfectly regular at
these points. It might seem at first sight that the signature of
$f_{\mu\nu}$ changes at these points.  However, this is not the
case since there exist two vectors $n_\mu=(1,0)$ and
$m_\mu=(\phi,a)$  such that $n^2=-1$, $m^2=1$ and $n\cdot m=0$
at every point either with $|\varphi|\leq 1$ or with
$|\varphi|\geq 1$.  Thus, the minisuperspace   $\Omega$ will
consist of all configurations $(a,\varphi)$ with $a>0$.  The
change of variables
\[
x=a,
\hspace{.5cm}\hspace{.5cm}
y=a\varphi
\]
transforms the metric in minisuperspace  into that of Minkowski
defined in the upper semiplane,  $x>0$, of Fig.~\ref{fig:min}.
The maximal analytic extension is obtained by extending the
range of $x$ to the whole real line so that the extended
minisuperspace $\bar\Omega$ will be the whole plane $ I\!\! R^2$
with the Minkowski metric, as happened in the case of minimal
coupling. Even more, the Wheeler--DeWitt operator, $\hat{\bar
H}$, in the extended minisuperspace  coincides with that of
minimal coupling.

The Wheeler--DeWitt operator $\hat H$ in the restricted
minisuperspace is not self--adjoint, although it is essentially
self--adjoint.  Suitable boundary conditions on the boundary
$x=0$ would allow us to choose the adecuate self--adjoint
extension. However, due to the hyperbolic character of the
Wheeler--DeWitt equation, the boundary conditions can only be
imposed on characteristic surfaces (actually, they are curves
since our minisuperspace  is two--dimensional) which, in this
model, are given by
\[
u=x+y \ ,
\hspace{.5cm}\hspace{.5cm}
v=x-y \ .
\]
Certainly, $x=0$ is not such a characteristic surface and,
therefore, it does not make any sense to try  to find solutions
that satisfy such conditions
\cite{pu55}. Since the zero volume three--geometry is
represented in this minisuperspace  by the configuration  $x=0$
with arbitrary $y$, it is not possible to give a precise meaning
to the regularity condition at  $a=0$ in the context of
restricted minisuperspace. It is, therefore, necessary to
consider its maximal analytic extension.

The quantization of this system is completly analogous to that
carried out in the case of minimal coupling and the expresions
given in that case are also valid here. The wormhole boundary
conditions reduce to the statement
\[
\Psi^{\rm o}(x,y)\in L^2( I\!\! R^2) \ .
\]
Once more, the regularity condition at zero volume
three--geometry is automatically satisfied, since the wave
function will be regular everywhere in the extended
minisuperspace   $\bar\Omega$ and, in particular, in the region
$x=0$.

The state  $\left|0,\varphi_0\right\rangle$, which was obtained
in Ref. \cite{ga91} as a path integral over all asymptotically
Euclidean metrics and over all matter fields whose asymptotic
configuration  $\varphi_0$ is such that the effective
gravitational coupling in this region  $\mbox{\small G}_{\rm
eff}=\mbox{\small G}/(1-\varphi_0^2)$ be positive, is
annihilated by the Wheeler--DeWitt operator in the extended
minisuperspace
\[
\hat{\bar H}\left|0,\varphi_0\right\rangle=0 \ ,
\]
that is,  $\Psi_{\varphi_0}^{\rm
o}(x,y)=\sigma^{-1}\left|0,\varphi_0\right\rangle$ satisfies the
Wheeler--DeWitt equation. In this example we can see the close
relation between the path integral and the canonical formalism.
In terms of the discrete basis states $\left|0,n\right\rangle$,
which are products of harmonic oscillator eigenfunctions as
first given in Ref. \cite{ha88}, this state can be written as
\[
\left|0,\varphi_0\right\rangle=
\sum_{n=0}^\infty\psi_n(\varphi_0)\left|0,n\right\rangle \ ,
\]
where
\[
\psi_n(\varphi_0)=\varphi_0^n\sqrt{1-\varphi_0^2} \ .
\]
The norm of this vector is
\[
\left\langle0,\varphi_0|0,\varphi_0\right\rangle=
\sum_{n=0}^\infty\varphi_0^{2n}(1-\varphi_0^2)=
\left\{\begin{array}{ll}1&{\rm if }\  |\varphi_0|<1\\
\infty&{\rm if }\ |\varphi_0|>1\end{array}\right. \ .
\]
Since the asymptotic region can be regarded as  classical and
observations can be made there, the asymptotic configurations
must agree with low energy physical predictions. In particular,
the effective gravitational coupling must be positive there,
i.e.  $\varphi_0^2<1$. Only in this situation the state
$\left|0,\varphi_0\right\rangle$ has any meaning \cite{ga91}.


\section{Summary and conclusions}
\label{sec:con}

In canonical quantum gravity,  wormhole wave functions satisfy
the Wheeler--DeWitt equation and the quantum constraints
associated to the invariance under spatial changes of
coordinates. Furthermore, they are subjected to suitable
boundary conditions which may be deduced from the path integral
formulation. We have restricted to minisuperspace, where only a
finite number of degrees of freedom have not been frozen out.
The asymptotically Euclidean character of wormholes makes the
wave function decrease exponentially for configurations that
represent arbitrarily large three--geometries. Likewise, the
wave function must  vanish for large values of the matter
fields. Also the wave function must be regular when the
three--volume vanishes since the spacetime manifold is not
singular at that point. The singularity of the three--geometry
is only due to the slicing procedure.

The boundary of minisuperspace consists of all those
configurations which are singular in some general sense,
including those mentioned above for which the minisuperspace
metric is singular although it corresponds to a coordinate
singularity. Therefore, it seems necessary consider the maximal
analytic extension of minisuperspace as the natural
configuration space for quantization. The true singularities,
such as those corresponding to infinite values of the
three--volume or the matter fields, belong to the boundary of
extended minisuperspace and the regular configurations,
including those associated to  zero three--volume, belong to its
interior.  Then, the wormhole boundary conditions can be simply
formulated in the following way: the wormhole wave functions
must be square integrable in the maximal analytic extension of
minisuperspace. This condition ensures that the wave function
vanishes at the truly singular configurations and guarantees its
regularity at any other configuration, including those which
represent zero volume three--geometries.  In fact, considering
the maximal analytic extension of minisuperspace, we avoid the
necessity  of imposing boundary conditions at zero volume
three--geometries which guarantee the self--adjointness of the
Wheeler--DeWitt operator. This operator is hyperbolic and,
therefore, in order to have a well posed boundary value problem,
boundary conditions should be imposed on characteristic surfaces
of this operator. However, the surface in minisuperspace
associated to vanishing three--geometries is not of this type,
in general, and consequently, it is meaningless to impose
boundary conditions on it.

Since wormholes are square integrable in the extended
minisuperspace, they form a Hilbert space whose inner product is
naturally induced by the minisuperspace metric, in which the
Wheeler--DeWitt operator is essentially self--adjoint. Then  we
can introduce an interpretation for the wormhole wave function
in terms of overlaps between different states. These overlaps
give an idea of the contribution of a given state to the
behaviour of a wormhole in another state. In particular, we can
conclude that giant wormholes should not contribute
significantly to any wormhole state.

The study of the whole spectrum of the Wheeler--DeWitt operator
is useful in the calculation of Green's functions in quantum
cosmology, since they can be written in terms of a complete set
of eigenfunctions of this operator. On the other hand, the
evaluation of the effective wormhole interaction beyond the
dilute wormhole approximation also requires  the whole spectrum
and not only the states that satisfy the Wheeler--DeWitt
equation, that is, the eigenstates of zero eigenvalue as happens
when this approximation is valid.

Finally, a consistent canonical quantum formulation of wormholes
requires a restriction of the gauge fixing conditions. For
instance, gauge conditions which vanish too fast with the
three--volume are not admissible. The canonical quantization
process contains an ambiguity, since different admissible gauge
fixing conditions give rise to different Hilbert spaces.
However physical results, such as effective interactions, are
independent of the gauge condition, due to the fact that they
can be given as path integrals.

\acknowledgements

I wish to thank Pedro Gonz\'alez--D\'{\i}az, Guillermo Mena
Marug\'an and Peter Tinyakov for their valuable comments on the
manuscript. I also thank   Instituto de Matem\'atica Aplicada y
F\'{\i}sica Fundamental, C.S.I.C., for hospitality. This work
was supported by DGCYT under contract PB91--0052 and by a Basque
Country Grant.



\begin{figure}
\caption{Maximal analytic extension of minisuperspace
corresponding to massless scalar field minimally or conformally
coupled to a Friedmann--\-Robertson--Walker spacetime.}
\label{fig:min}
\end{figure}

\begin{figure}
\caption{Behaviour of the modified Bessel function
$K_{i\frac{k}{2}}(a^2/2)$. The value $a=\protect\sqrt{|k|}$
represents the wormhole throat radius.}
\label{fig:bes}
\end{figure}

\end{document}